\documentclass[envcountsame]{llncs}

\usepackage{t1enc} 
\usepackage{amsmath}
\usepackage{latexsym}
\usepackage{theorem}
\usepackage{amssymb}
\usepackage{url}

\newcommand{\I}{}

\newcommand{\ifshort}[1]{\ifthenelse{\equal{\version}{short}}{#1}{}}
\newcommand{\iflong}[1]{\ifthenelse{\equal{\version}{long}}{#1}{}}

\newcommand{\comment}[1]{}
\newcommand{\bu}{$\bullet$}

\newcommand{\vide}{\emptyset}

\newcommand{\dom}{\mr{dom}}

\newcommand{\FV}{\mr{FV}}
\newcommand{\pos}{\mr{Pos}}

\renewcommand{\a}{\rightarrow}
\newcommand{\A}{\Rightarrow}

\newcommand{\ad}{\downarrow}

\renewcommand{\to}{\mapsto}

\newcommand{\ab}{\a_\b}

\newcommand{\ar}{\a_\cR}
\newcommand{\abr}{\a_{\b\cR}}

\renewcommand{\I}[1]{[\![#1]\!]}

\newcommand{\all}{\forall}
\newcommand{\ou}{\vee}

\newcommand{\et}{\wedge}
\newcommand{\non}{\neg}
\newcommand{\st}{\star}
\newcommand{\B}{\Box} 
\renewcommand{\th}{\vdash}

\newcommand{\sle}{\subseteq}

\newcommand{\tgt}{\rhd}

\newcommand{\mi}{\mathit}
\newcommand{\mc}{\mathcal}

\newcommand{\mr}{\mathrm}

\renewcommand{\b}{\beta}
\newcommand{\g}{\gamma}
\newcommand{\G}{\Gamma}
\renewcommand{\d}{\delta}
\newcommand{\D}{\Delta}
\newcommand{\ep}{\epsilon}
\newcommand{\vep}{\varepsilon}

\renewcommand{\t}{\theta}

\newcommand{\io}{\iota}
\newcommand{\ka}{\kappa}
\newcommand{\la}{\lambda}

\renewcommand{\r}{\rho}
\newcommand{\s}{\sigma}

\newcommand{\vphi}{\varphi}
\newcommand{\w}{\omega}

\newcommand{\cF}{\mc{F}}
\newcommand{\cG}{\mc{G}}

\newcommand{\cN}{\mc{N}}

\newcommand{\cR}{\mc{R}}
\newcommand{\cS}{\mc{S}}
\newcommand{\cT}{\mc{T}}

\newcommand{\cX}{\mc{X}}

\newcommand{\va}{{\vec{a}}}

\newcommand{\vc}{{\vec{c}}}

\newcommand{\vl}{{\vec{l}}}

\newcommand{\vq}{{\vec{q}}}

\newcommand{\vt}{{\vec{t}}}
\newcommand{\vu}{{\vec{u}}}
\newcommand{\vv}{{\vec{v}}}
\newcommand{\vw}{{\vec{w}}}
\newcommand{\vx}{{\vec{x}}}
\newcommand{\vy}{{\vec{y}}}
\newcommand{\vz}{{\vec{z}}}

\newcommand{\vA}{{\vec{A}}}

\newcommand{\vQ}{{\vec{Q}}}
\newcommand{\vR}{{\vec{R}}}
\newcommand{\vS}{{\vec{S}}}
\newcommand{\vT}{{\vec{T}}}
\newcommand{\vU}{{\vec{U}}}
\newcommand{\vV}{{\vec{V}}}
\newcommand{\vW}{{\vec{W}}}

\newcommand{\xS}{\{x\to S\}}

\newcommand{\xu}{\{x\to u\}}

\newcommand{\vxl}{\{\vx\to\vl\}}

\newcommand{\vxu}{\{\vx\to\vu\}}

\newcommand{\vyu}{\{\vy\to\vu\}}

\newenvironment{rul}%
  {$\begin{array}{rcl}}%
  {\end{array}$}
  {\begin{center}\begin{rul}}%
  {\end{rul}\end{center}}

\newenvironment{rew}[1][~~\a~~]%
  {$\begin{array}{r@{#1}l}}%
  {\end{array}$}
\newenvironment{rewc}[1][~~\a~~]%
  {\begin{center}\begin{rew}[#1]}%
  {\end{rew}\end{center}}

\newenvironment{typc}[1][\,:\,]{\begin{rewc}[#1]}{\end{rewc}}

\newcounter{counter}

{\theorembodyfont{\rmfamily} 
  \newtheorem{dfn}[counter]{Definition}

}

\newenvironment{lstgeneric}[2]
  {\begin{list}{#1}{\topsep=.5mm\itemsep=.5mm\parsep=0mm%
    \itemindent=-3ex\labelsep=1ex\labelwidth=0ex #2}}
  {\end{list}}

\newenvironment{lst}[1]
  {\begin{lstgeneric}{#1}{\itemindent=-1ex}}
  {\end{lstgeneric}}

\newenvironment{enumi}[1]
  {\begin{lstgeneric}{}{\usecounter{enumi}\leftmargin=7mm%
    }}
  {\end{lstgeneric}}

\newenvironment{bfenumii}[1]
  {\begin{lstgeneric}{}{\usecounter{enumii}\leftmargin=7mm%
    }}
  {\end{lstgeneric}}

\newcommand{\tf}{{\tau_f}}
\newcommand{\tg}{{\tau_g}}

\newcommand{\tC}{{\tau_C}}

\newcommand{\XB}{\cX^\B}

\newcommand{\SN}{\cS\cN}

\newcommand{\at}{\alpha}

\begin{document}


\title{Inductive types in the\\
Calculus of Algebraic Constructions}

\author{Fr\'ed\'eric Blanqui}

\institute{Laboratoire d'Informatique de l'\'Ecole Polytechnique\\
91128 Palaiseau Cedex, France\\
\email{blanqui@lix.polytechnique.fr}}

\maketitle

\begin{abstract}
In a previous work, we proved that almost all of the Calculus of
Inductive Constructions (CIC), the basis of the proof assistant Coq,
can be seen as a Calculus of Algebraic Constructions (CAC), an
extension of the Calculus of Constructions with functions and
predicates defined by higher-order rewrite rules. In this paper, we
prove that CIC as a whole can be seen as a CAC, and that it can be
extended with non-strictly positive types and inductive-recursive
types together with non-free constructors and pattern-matching on
defined symbols.
\end{abstract}


\section{Introduction}
\label{sec-intro}

There has been different proposals for defining inductive types and
functions in typed systems. In Girard's polymorphic $\la$-calculus or
in the Calculus of Constructions (CC) \cite{coquand88ic}, data types
and functions can be formalized by using impredicative encodings,
difficult to use in practice, and computations are done by
$\b$-reduction only. In Martin-L\"of's type theory or in the Calculus
of Inductive Constructions (CIC) \cite{coquand88colog}, inductive
types and their induction principles are first-class objects,
functions can be defined by induction and computations are done by
$\io$-reduction. For instance, for the type $nat$ of natural numbers,
the recursor $rec:(P:nat\A\st)(u:P0)(v:(n:nat)Pn\A P(sn))(n:nat)Pn$ is
defined by the following $\io$-rules:

\begin{rewc}[~~\a_\io~~]
rec~P~u~v~0 & u\\
rec~P~u~v~(s~n) & v~n~(rec~P~u~v~n)\\
\end{rewc}

Finally, in the algebraic setting \cite{dershowitz90book}, functions
are defined by using rewrite rules and computations are done by
applying these rules. Since both $\b$-reduction and $\io$-reduction
are particular cases of higher-order rewriting \cite{klop93tcs},
proposals soon appeared for integrating all these approaches. Starting
with \cite{jouannaud91lics,barbanera94lics}, this objective culminated
with \cite{blanqui01lics,blanqui01thesis,blanqui03cac} in which almost
all of CIC can be seen as a Calculus of Algebraic Constructions (CAC),
an extension of CC with functions and predicates defined by
higher-order rewrite rules. In this paper, we go one step further in
this direction and capture all previous proposals, and much more.


Let us see the two examples of recursors that are allowed in CIC but
not in CAC \cite{paulin01pc}. The first example is a third-order
definition of finite sets of natural numbers (represented as
predicates over $nat$):

\newcommand{\fin}{{\mi{fin}}}
\newcommand{\femp}{\mi{femp}}
\newcommand{\fadd}{\mi{fadd}}

\begin{typc}[\,]
\fin: & (nat\A\st)\A\st\\
\femp: & \fin\,\vide\\
\fadd: & (x:nat)(p:nat\A\st)\fin\, p\A \fin(\mr{add}~x~p)\\
rec: & (Q:(nat\A\st)\A\st)Q\vide\\
& \A ((x:nat)(p:nat\A\st)\fin\, p\A Qp\A Q(\mr{add}~x~p))\\
& \A (p:nat\A\st)\fin\, p\A Qp\\
\end{typc}

\noindent
where $\vide= [y:nat]\bot$ represents the empty set, $\mr{add}~x~p=
[y:nat]y=x\ou (p~y)$ represents the set $\{x\}\cup p$, and the weak
recursor $rec$ (recursor for defining objects) is defined by the
rules:

\begin{rewc}
rec~Q~u~v~p'~\femp & u\\
rec~Q~u~v~p'~(\fadd~x~p~h) & v~x~p~h~(rec~Q~u~v~p~h)\\
\end{rewc}

The problem comes from the fact that, in $\fin(\mr{add}~x~p)$, the
output type of $\fadd$, the predicate $p$ is not a parameter of
$\fin$.\footnote{This is also the reason why the corresponding strong
recursor, that is, the recursor for defining types or predicates, is
not allowed in CIC ($p$ could be ``bigger'' than $\fin$).} This can be
generalized to any big/impredicative dependent type, that is, to any
type having a constructor with a predicate argument which is not a
parameter. Formally, if $C:(\vz:\vV)\st$ is a type and
$c:(\vx:\vT)C\vv$ is a constructor of $C$ then, for all predicate
variable $x$ occurring in some $T_j$, there must be some argument
$v_{\io_x}=x$, a condition called (I6) in \cite{blanqui01thesis}.


\newcommand{\JMeq}{\mi{JMeq}}
\newcommand{\refl}{\mi{refl}}

The second example is John Major's equality which is intended to equal
terms of different types \cite{mcbride99thesis}:

\begin{typc}[\,]
\JMeq: & (A:\st)A\A(B:\st)B\A\st\\
\refl: & (A:\st)(x:A)(\JMeq~A~x~A~x)\\
rec: & (A:\st)(x:A)(P:(B:\st)B\A\st)(P~A~x)\\
& \A(B:\st)(y:B)(\JMeq~A~x~B~y)\A (P~B~y)\\
\end{typc}

\noindent
where $rec$ is defined by the rule:

\begin{rewc}
rec~C~x~P~h~C~x~(\refl~C~x) & h\\
\end{rewc}

\noindent
Here, the problem comes from the fact that the argument for $B$ is
equal to the argument for $A$. This can be generalized to any
polymorphic type having a constructor with two equal type
parameters. From a rewriting point of view, this is like having
pattern-matching or non-linearities on predicate arguments, which is
known to create inconsistencies in some cases
\cite{harper99ipl}. Formally, a rule $f\vl\a r$ with $f:(\vx:\vT)U$ is
{\em safe} if, for all predicate argument $x_i$, $l_i$ is a variable
and, if $x_i$ and $x_j$ are two distinct predicate arguments, then
$l_i\neq l_j$. An inductive type is {\em safe} if the corresponding
$\io$-rules are safe.


By using what is called in Matthes' terminology \cite{matthes98thesis}
an {\em elimination-based} interpretation instead of the {\em
introduction-based} interpretation that we used in
\cite{blanqui01thesis}, we prove that recursors for types like $\fin$
or $\JMeq$ can be accepted, hence that CAC essentially subsumes
CIC. In addition, we prove that it can be extended to non-strictly
positive types (Section \ref{sec-pos}) and to inductive-recursive
types \cite{dybjer00jsl} (Section \ref{sec-indrec}).




\section{The Calculus of Algebraic Constructions (CAC)}

We assume the reader familiar with typed $\la$-calculi
\cite{barendregt92book} and rewriting \cite{dershowitz90book}. The
Calculus of Algebraic Constructions (CAC) \cite{blanqui01thesis}
simply extends CC by considering a set $\cF$ of {\em symbols},
equipped with a total quasi-ordering $\ge$ (precedence) whose strict
part is well-founded, and a set $\cR$ of {\em rewrite rules}. The
terms of CAC are:

\begin{center}
$t ::= s ~|~ x ~|~ f ~|~ [x:t]u ~|~ tu ~|~ (x:t)u$
\end{center}

\noindent
where $s\in\cS=\{\st,\B\}$ is a {\em sort}, $x\in\cX$ a {\em
variable}, $f\in\cF$, $[x:t]u$ an {\em abstraction}, $tu$ an {\em
application}, and $(x:t)u$ a {\em dependent product}, written $t\A u$
if $x$ does not freely occur in $u$. We denote by $\FV(t)$ the set of
free variables of $t$, by $\pos(t)$ the set of Dewey's positions of
$t$, and by $\dom(\t)$ the {\em domain} of a substitution $\t$.

The sort $\st$ denotes the universe of types and propositions, and the
sort $\B$ denotes the universe of predicate types (also called {\em
kinds}). For instance, the type $nat$ of natural numbers is of type
$\st$, $\st$ itself is of type $\B$ and $nat\A\st$, the type of
predicates over $nat$, is of type $\B$.

Every symbol $f$ is equipped with a sort $s_f$, an {\em arity} $\at_f$
and a type $\tf$ which may be any closed term of the form $(\vx:\vT)U$
with $|\vx|=\at_f$ ($|\vx|$ is the length of $\vx$). We denote by
$\G_f$ the environment $\vx:\vT$. The terms only built from variables
and applications of the form $f\vt$ with $|\vt|=\at_f$ are called {\em
algebraic}.


A rule for typing symbols is added to the typing rules of CC:

\begin{center}
(symb) \quad $\cfrac{\th\tf:s_f}{\th f:\tf}$
\end{center}

A {\em rewrite rule} is a pair $l\a r$ such that (1) $l$ is algebraic,
(2) $l$ is not a variable, and (3) $\FV(r)\sle\FV(l)$. A symbol $f$
with no rule of the form $f\vl\a r$ is {\em constant}, otherwise it is
(partially) {\em defined}. We also assume that, in every rule $f\vl\a
r$, the symbols occurring in $r$ are smaller than or equivalent to
$f$.

\newcommand{\br}{{\b\cR}}

Finally, in CAC, $\br$-equivalent types are identified. More
precisely, in the type conversion rule of CC, $\ad_\b$ is replaced by
$\ad_\br$:

\begin{center}
(conv) \quad $\cfrac{\G\th t:T \quad T\ad_\br T' \quad \G\th
T':s}{\G\th t:T'}$
\end{center}

\noindent
where $u\ad_\br v$ iff there exists a term $w$ such that $u\abr^* w$
and $v\abr^* w$, $\abr^*$ being the reflexive and transitive closure
of $\a=\ab\cup\ar$. This rule means that any term $t$ of type $T$ in
the environment $\G$ is also of type $T'$ if $T$ and $T'$ have a
common reduct (and $T'$ is of type some sort $s$). For instance, if
$t$ is a proof of $P(2+2)$ then $t$ is also a proof of $P(4)$ if $\cR$
contains the following rules:

\begin{rewc}
x+0 & x\\
x+(s~y) & s~(x+y)\\
\end{rewc}

This allows to decrease the size of proofs by an important factor, and
to increase the automation as well. {\bf All over the paper, we assume
that $\a$ is confluent}.

A substitution $\t$ {\em preserves typing from $\G$ to $\D$}, written
$\t:\G\leadsto\D$, if, for all $x\in\dom(\G)$, $\D\th x\t:x\G\t$,
where $x\G$ is the type associated to $x$ in $\G$. Type-preserving
substitutions enjoy the following important property: if $\G\th t:T$
and $\t:\G\leadsto\D$ then $\D\th t\t:T\t$.


For ensuring the {\em subject reduction} property (preservation of
typing under reduction), every rule $f\vl\a r$ is equipped with an
environment $\G$ and a substitution $\r$ such that, if $f:(\vx:\vT)U$
and $\g=\vxl$, then $\G\th f\vl\r:U\g\r$ and $\G\th r:U\g\r$. The
substitution $\r$ allows to eliminate non-linearities due to
typing. For instance, the concatenation on polymorphic lists (type
$list:\st\A\st$ with constructors $nil:(A:\st)listA$ and
$cons:(A:\st)A\A listA\A listA$) of type $(A:\st)listA\A listA\A
listA$ can be defined by:

\begin{rewc}
app~A~(nil~A')~l' & l'\\
app~A~(cons~A'~x~l)~l' & cons~A~x~(app~A~x~l~l')\\
app~A~(app~A'~l~l')~l'' & app~A~l~(app~A~l'~l'')\\
\end{rewc}

\noindent
with $\G=A:\st,x:A,l:listA,l':listA$ and $\r=\{A'\to A\}$. For
instance, $app~A~(nil~A')$ is not typable in $\G$ (since
$A'\notin\dom(\G)$) but becomes typable if we apply $\r$. This does
not matter since, if an instance $app~A\s~(nil~A'\s)$ is typable then
$A\s$ is convertible to $A'\s$. Eliminating non-linearities makes
rewriting more efficient and the proof of confluence easier.




\section{Strong normalization}
\label{sec-sn}

Typed $\la$-calculi are generally proved strongly normalizing by using
Tait and Girard's technique of {\em reducibility candidates}
\cite{girard88book}. The idea of Tait, later extended by Girard to the
polymorphic $\la$-calculus, is to strengthen the induction
hypothesis. Instead of proving that every term is strongly
normalizable (set $\SN$), one associates to every type $T$ a set
$\I{T}\sle\SN$, the {\em interpretation} of $T$, and proves that every
term $t$ of type $T$ is {\em computable}, {\em i.e.}  belongs to
$\I{T}$. Hereafter, we follow the proof given in \cite{blanqui03short}
which greatly simplifies the one given in \cite{blanqui01thesis}.


\begin{dfn}[Reducibility candidates]
A term $t$ is {\em neutral} if it is not an abstraction, not of the
form $c\vt$ with $c:(\vy:\vU)C\vv$ and $C$ constant, nor of the form
$f\vt$ with $f$ defined and $|\vt|<\at_f$. We inductively define the
complete lattice $\cR_t$ of the interpretations for the terms of type
$t$, the ordering $\le_t$ on $\cR_t$, and the greatest element
$\top_t\in\cR_t$ as follows.

\begin{lst}{--}
\item $\cR_t=\{\vide\}$, $\le_t=\sle$ and $\top_t=\vide$ if $t\neq\B$
and $\G\not\th t:\B$.

\item $\cR_s$ is the set of subsets $R\sle\cT$ such that:
\begin{bfenumii}{R}
\item $R\sle\SN$ (strong normalization).
\item If $t\in R$ then $\a\!\!(t)=\{t'~|~t\a t'\}\sle R$ (stability by
reduction).
\item If $t$ is neutral and $\a\!\!(t)\sle R$ then $t\in R$ (neutral terms).
\end{bfenumii}
Furthermore, $\le_s=\sle$ and $\top_s=\SN$.

\item $\cR_{(x:U)K}$ is the set of functions $R$ from $\cT\times\cR_U$
to $\cR_K$ such that $R(u,S)=R(u',S)$ whenever $u\a u'$,
$R\le_{(x:U)K} R'$ iff, for all $(u,S)\in\cT\times\cR_U$, $R(u,S)\le_K
R'(u,S)$, and $\top_{(x:U)K}(u,S)=\top_K$.
\end{lst}
\end{dfn}

Note that $\cR_t=\cR_{t'}$ whenever $t\a t'$ and that, for all
$R\in\cR_s$, $\cX\sle R$.


\begin{dfn}[Interpretation schema]
\label{def-schema-int}

A {\em candidate assignment} is a function $\xi$ from $\cX$ to
$\bigcup \,\{\cR_t ~|~ t\in\cT\}$. An assignment $\xi$ {\em validates}
an environment $\G$, written $\xi\models\G$, if, for all
$x\in\dom(\G)$, $x\xi\in \cR_{x\G}$. An {\em interpretation} for a
symbol $f$ is an element of $\cR_\tf$. An {\em interpretation} for a
set $\cG$ of symbols is a function which, to each symbol $g\in\cG$,
associates an interpretation for $g$.

The {\em interpretation} of $t$ w.r.t. a candidate assignment $\xi$,
an interpretation $I$ for $\cF$ and a substitution $\t$, is defined by
induction on $t$ as follows.

\begin{lst}{\bu}
\item $\I{t}^I_{\xi,\t}= \top_t$ if $t$ is an object or a sort,
\item $\I{x}^I_{\xi,\t}= x\xi$,
\item $\I{f}^I_{\xi,\t}= I_f$,
\item $\I{(x:U)V}^I_{\xi,\t}= \{t\in\cT~|~ \all u\in\I{U}^I_{\xi,\t},
\all S\in\cR_U, tu\in\I{V}^I_{\xi_x^S,\t_x^u}\}$,
\item $\I{[x:U]v}^I_{\xi,\t}(u,S)= \I{v}^I_{\xi_x^S,\t_x^u}$,
\item $\I{tu}^I_{\xi,\t}= \I{t}^I_{\xi,\t}(u\t,\I{u}^I_{\xi,\t})$,
\end{lst}

\noindent
where $\xi_x^S=\xi\cup\xS$ and $\t_x^u=\t\cup\xu$. A substitution $\t$
is {\em adapted} to a $\G$-assignment $\xi$ if $\dom(\t)\sle\dom(\G)$
and, for all $x\in\dom(\t)$, $x\t\in \I{x\G}^I_{\xi,\t}$. A pair
$(\xi,\t)$ is {\em $\G$-valid}, written $\xi,\t\models\G$, if
$\xi\models\G$ and $\t$ is adapted to $\xi$.
\end{dfn}


Note that $\I{t}_{\xi,\t}^I= \I{t}_{\xi',\t'}^{I'}$ whenever $\xi$ and
$\xi'$ agree on the predicate variables free in $t$, $\t$ and $\t'$
agree on the variables free in $t$, and $I$ and $I'$ agree on the
symbols occurring in $t$. The difficult point is then to define an
interpretation for predicate symbols and to prove that every symbol
$f$ is computable ({\em i.e.} $f\in\I{\tf}$).


Following previous works on inductive types
\cite{mendler87thesis,werner94thesis}, the interpretation of a
constant predicate symbol $C$ is defined as the least fixpoint of a
monotone function $I\mapsto\vphi_C^I$ on the complete lattice
$\cR_\tC$. Following Matthes \cite{matthes98thesis}, there is
essentially two possible definitions that we illustrate by the case of
$nat$. The {\em introduction-based} definition:

\begin{center}
$\vphi_{nat}^I= \{t\in\SN~|~ t\a^* su\A u\in I\}$
\end{center}

\noindent
and the {\em elimination-based} definition:

\begin{center}
$\vphi_{nat}^I= \{t\in\cT~|~ \all (\xi,\t)\, \G\mbox{-valid},\,
rec~P\t~u\t~v\t~t\in\I{Pn}^I_{\xi,\t_n^t}\}$
\end{center}

\noindent
where $\G=P:nat\A\st,u:P0,v:(n:nat)Pn\A P(sn)$. In both cases, the
monotonicity of $\vphi_{nat}$ is ensured by the fact that $nat$ occurs
only positively\footnote{$X$ occurs positively in $Y\A X$ and
negatively in $X\A Y$. In Section \ref{sec-indrec}, we give an
extended definition of positivity for dealing with inductive-recursive
types \cite{dybjer00jsl}.} in the types of the arguments of its
constructors, a common condition for inductive types.\footnote{Mendler
proved that recursors for negative types are not normalizing
\cite{mendler87thesis}. Take for instance an inductive type $C$ with a
constructor $c:(C\a nat)\a C$. Assume now that we have $p:C\a (C\a
nat)$ defined by the rule $p(cx)\a x$ (case analysis). Then, by taking
$\w= [x:C](px)x$, we get $\w(c\w)\ab p(c\w)(c\w)\a \w(c\w)\ab
\ldots$}.

In \cite{blanqui01thesis}, we used the introduction-based approach
since this allows us to have non-free constructors and
pattern-matching on defined symbols, which is forbidden in CIC and
does not seem possible with the elimination-based approach. Indeed, in
CAC, it is possible to formalize the type $int$ of integers by taking
the symbols $0:int$, $s:int\A int$ and $p:int\A int$, together with
the rules:

\begin{rewc}
s~(p~x) & x\\
p~(s~x) & x\\
\end{rewc}

It is also possible to have the following rule on natural numbers:

\begin{rewc}
x\times(y+z) & (x\times y)+(x\times z)\\
\end{rewc}

\newcommand{\acc}{\mr{Acc}}

To this end, we extended the notion of constructor by considering as
{\em constructor} any symbol $c$ whose output type is a constant
predicate symbol $C$ (perhaps applied to some arguments). Then, the
arguments of $c$ that can be used to define the result of a function
are restricted to the arguments, called {\em accessible}, in the type
of which $C$ occurs only positively. We denote by $\acc(c)$ the set of
accessible arguments of $c$. For instance, $x$ is accessible in $sx$
since $nat$ occurs only positively in the type of $x$. But, we also
have $x$ and $y$ accessible in $x+y$ since $nat$ occurs only
positively in the types of $x$ and $y$. So, $+$ can be seen as a
constructor too.

With this approach, we can safely take:

\begin{center}
$\vphi_{nat}^I= \{t\in\SN~|~ \all f, t\a^* f\vu\A \all j\in \acc(f),
u_j\in \I{U_j}^I_{\xi,\t}\}$
\end{center}

\noindent
where $f:(\vy:\vU)C\vv$ and $\t=\vyu$, whenever an appropriate
assignment $\xi$ for the predicate variables of $U_j$ can be defined,
which is possible only if the condition (I6) is satisfied (see the
type $\fin$ in Section \ref{sec-intro}).




\section{Extended recursors}

As we introduced an extended notion of constructor for dealing with
the intro\-duc\-tion-based method, we now introduce an extended notion
of recursor for dealing with the elimination-based method.

\newcommand{\rec}{\cR ec}

\begin{definition}[Extended recursors]
\label{def-rec}
A {\em pre-recursor} for a constant predicate symbol $C:(\vz:\vV)\st$
is any symbol $f$ such that:

\begin{lst}{\bu}
\item the type of $f$ is of the form\footnote{Our examples may not
always fit in this form but since, in an environment, two types that
do not depend on each other can be permuted, this does not matter.}
$(\vz:\vV)(z:C\vz)W$,
\item every rule defining $f$ is of the form $f\vz t\vu\a r$ with
  $\FV(r)\cap \{\vz\}=\vide$,
\item $f\vv t\vu$ is head-reducible only if $t$ is constructor-headed.
\end{lst}

\noindent
A pre-recursor $f$ is a {\em recursor} if it satisfies the following
{\em positivity conditions}:\footnote{In Section \ref{sec-indrec}, we
give weaker conditions for dealing with inductive-recursive types.}

\begin{lst}{\bu}
\item no constant predicate $D>C$ or defined predicate $F$ occurs in $W$,
\item every constant predicate $D\simeq C$ occurs only positively in
  $W$.
\end{lst}

\noindent
A recursor of sort $\st$ (resp. $\B$) is {\em weak} (resp. {\em
strong}). Finally, we assume that every type $C$ has a set $\rec(C)$
(possibly empty) of recursors.
\end{definition}


For the types $C$ whose set of recursors $\rec(C)$ is not empty, we
define the interpretation of $C$ with the elimination-based method as
follows. For the other types, we keep the introduction-based method.

\begin{definition}[Interpretation of inductive types]
\label{def-int}
If every $t_i$ has a normal form $t_i^*$ then $\vphi^I_C(\vt,\vS)$ is
the set of terms $t$ such that, for all $f\in\rec(C)$ of type
$(\vz:\vV)(z:C\vz)(\vy:\vU)V$, $\vy\xi$ and $\vy\t$, if
$\xi_\vz^\vS,\t_\vz^\vt{}_z^t\models\vy:\vU$ then $f\vt^* t\vy\t\in
\I{V}_{\xi_\vz^\vS,\t_\vz^\vt{}_z^t}^I$. Otherwise,
$\vphi^I_C(\vt,\vS)=\SN$.
\end{definition}

The fact that $\vphi$ is monotone, hence has a least fixpoint, follows
from the positivity conditions. One can easily check that $\vphi^I_C$
is stable by reduction: if $\vt\a\vt'$ then
$\vphi^I_C(\vt,\vS)=\vphi^I_C(\vt',\vS)$. We now prove that
$\vphi^I_C(\vt,\vS)$ is a candidate.


\begin{lemma}
$\vphi^I_C(\vt,\vS)$ is a candidate.
\end{lemma}

\begin{proof}
\begin{enumi}{R}
\item Let $t\in R$. We must prove that $t\in\SN$. Since
  $\rec(C)\neq\vide$, there is at least one recursor $f$. Take
  $y_i\t=y_i$ and $y_i\xi=\top_{U_i}$. We clearly have
  $\xi_\vz^\vS,\t_\vz^\vt{}_z^t\models\vy:\vU$. Therefore, $f\vt^*
  t\vy\in S=\I{V}_{\xi_\vz^\vS,\t_\vz^\vt{}_z^t}^I$. Now, since $S$
  satisfies (R1), $f\vt^*t\vy\in\SN$ and $t\in\SN$.

\item Let $t\in R$ and $t'\in\,\a\!\!(t)$. We must prove that $t'\in
  R$, hence that $f\vt^*t'\vy\t\in
  S=\I{V}_{\xi_\vz^\vS,\t_\vz^\vt{}_z^t}^I$. This follows from the
  fact that $f\vt^*t\vy\t\in S$ (since $t\in R$) and $S$ satisfies
  (R2).

\item Let $t$ be a neutral term such that $\a\!\!(t)\sle R$. We must
  prove that $t\in R$, hence that $u=f\vt^* t\vy\t\in
  S=\I{V}_{\xi_\vz^\vS,\t_\vz^\vt{}_z^t}^I$. Since $u$ is neutral and
  $S$ satisfies (R3), it suffices to prove that $\a\!\!(u)\sle
  S$. Since $\vy\t\in\SN$ by (R1), we proceed by induction on $\vy\t$
  with $\a$ as well-founded ordering. The only difficult case could be
  when $u$ is head-reducible, but this is not possible since $t$ is
  neutral, hence not constructor-headed.\qed
\end{enumi}
\end{proof}




\section{Admissible recursors}
\label{sec-admis}


Since we changed the interpretation of constant predicate symbols, we
must check several things in order to preserve the strong
normalization result of \cite{blanqui01thesis}.

\begin{lst}{\bu}
\item We must make sure that the interpretation of primitive types is
still $\SN$ since this is used for proving the computability of
first-order symbols and the interpretation of some defined predicate
symbols (see Lemma \ref{lem-prim}).

\item We must also prove that every symbol is computable.

\begin{lst}{--}
\item For extended recursors, this follows from the definition of the
interpretation for constant predicate symbols, and thus, does not
require safety.

\item For first-order symbols, nothing is changed.

\item For higher-order symbols distinct from recursors, we must make
sure that the accessible arguments of a computable constructor-headed
term are computable.

\item For constructors, this does not follow from the interpretation
for constant predicate symbols anymore. We therefore have to prove it.
\end{lst}
\end{lst}


\noindent
We now define general conditions for these requirements to be
satisfied.

\begin{definition}[Admissible recursors]
\hfill Assume that every constructor is\\ equipped with a set
$\acc(c)\sle\{1,\ldots,\at_c\}$ of {\em accessible arguments}. Let
$C:(\vz:\vV)\st$ be a constant predicate symbol. $\rec(C)$ is {\em
complete w.r.t. accessibility} if, for all $c:(\vx:\vT)C\vv$,
$j\in\acc(c)$, $\vx\eta$ and $\vx\s$, if $\eta\models\G_c$,
$\vv\s\in\SN$ and $c\vx\s\in \I{C\vv}_{\eta,\s}$ then $x_j\s\in
\I{T_j}_{\eta,\s}$.

A recursor $f:(\vz:\vV)(z:C\vz)(\vy:\vU)V$ is {\em head-computable
w.r.t} a constructor $c:(\vx:\vT)C\vv$ if, for all $\vx\eta$, $\vx\s$,
$\vy\xi$, $\vy\t$, $\vS=\I{\vv}_{\eta,\s}$, if $\eta,\s\models\G_c$
and $\xi_\vz^\vS,\t_\vz^{\vv\s}{}_z^{c\vx\s}\models\vy:\vU$, then
every head-reduct of $f\vv\s(c\vx\s)\vy\t$ belongs to
$\I{V}_{\xi_\vz^\vS,\t_\vz^{\vv\s}{}_z^{c\vx\s}}$. A recursor is {\em
head-computable} if it is head-computable w.r.t. every
constructor. $\rec(C)$ is {\em head-computable} if all its recursors
are head-computable.

$\rec(C)$ is {\em admissible} if it is head-computable and complete
w.r.t. accessibility.
\end{definition}


We first prove that the interpretation of primitive types is $\SN$.

\begin{lemma}[Primitive types]
\label{lem-prim}
Types equivalent to $C$ are {\em primitive} if, for all $D\simeq C$,
$D:\st$ and, for all $d:(\vx:\vT)D$, $\acc(d)=\{1,\ldots,\at_d\}$ and
every $T_j$ is a primitive type $E\le C$. Let $C:\st$ be a primitive
symbol. If recursors are head-computable then $I_C=\SN$.
\end{lemma}

\begin{proof}
By definition, $I_C\sle\SN$. We prove that, if $t\in\SN$ then $t\in
I_C$, by induction on $t$ with $\a\cup\,\tgt$ as well-founded
ordering. Let $f:(z:C)(\vy:\vU)V$ be a recursor, $\vy\xi$ and $\vy\t$
such that $\xi,\t_z^t\models \vy:\vU$. We must prove that
$v=ft\vy\t\in S=\I{V}_{\xi,\t_z^t}$. Since $v$ is neutral, it suffices
to prove that $\a\!\!(v)\sle S$. We proceed by induction on $t\vy\t$
with $\a$ as well-founded ordering ($\vy\t\in\SN$ by R1). If the
reduction takes place in $t\vy\t$, we can conclude by induction
hypothesis. Assume now that $v'$ is a head-reduct of $v$. By
assumption on recursors (Definition \ref{def-rec}), $t$ is of the form
$c\vu$ with $c:(\vx:\vT)C$. Since $C$ is primitive, every $u_j$ is
accessible and every $T_j$ is a primitive type $D\le C$. By induction
hypothesis, $u_j\in I_D$. Therefore, $\vide,\vxu\models\G_c$ and,
since $\xi,\t_z^t\models \vy:\vU$ and recursors are head-computable,
$v'\in S$.\qed
\end{proof}


\begin{theorem}[Strong normalization]
\label{thm-admis}
Assume that every constant predicate symbol $C$ is equipped with an
admissible set $\rec(C)$ of extended recursors distinct from
constructors. If $\a$ is confluent and strong recursors and symbols
that are not recursors satisfy the conditions given in
\cite{blanqui01thesis} then $\b\cup\cR$ is strongly normalizing.
\end{theorem}

\begin{proof}
Let $\th_f$ (resp. $\th_f^<$) be the typing relation of the CAC whose
symbols are (resp. strictly) smaller than $f$. By induction on $f$, we
prove that, if $\G\th_f t:T$ and $\xi,\t\models\G$ then $t\t\in
\I{T}_{\xi,\t}$. By (symb), if $g\le f$ and $\th_f g:\tg$ then
$\th_f^< \tg:s_g$. Therefore, the induction hypothesis can be applied
to the subterms of $\tg$.

We first prove that recursors are computable. Let
$f:(\vz:\vV)(z:C\vz)(\vy:\vU)V$ be a recursor and assume that
$\xi,\t\models\G_f$. We must prove that $v=f\vz\t z\t\vy\t\in
S=\I{V}_{\xi,\t}$. Since $v$ is neutral, it suffices to prove that
$\a\!\!(v)\sle S$. We proceed by induction on $\vz\t z\t\vy\t$ with
$\a$ as well-founded ordering ($\vz\t z\t\vy\t\in\SN$ by R1). If the
reduction takes place in $\vz\t z\t\vy\t$, we conclude by induction
hypothesis. Assume now that we have a head-reduct $v'$. By assumption
on recursors (Definition \ref{def-rec}), $z\t$ is of the form $c\vu$
with $c:(\vx:\vT)C\vv$, and $v'$ is a head-reduct of $v_0=
f\vz\t^*z\t\vy\t$ where $\vz\t^*$ are the normal forms of
$\vz\t$. Since $\xi,\t\models\G_f$, we have $z\t=c\vu\in
\I{C\vz}_{\xi,\t}= I_C(\vz\t,\vz\xi)$. Therefore, $v_0\in S$ and, by
(R2), $v'\in S$.

We now prove that constructors are computable. Let $c:(\vx:\vT)C\vv$
be a constructor of $C:(\vz:\vV)\st$, $\vx\eta$ and $\vx\s$ such that
$\eta,\s\models\G_c$. We must prove that $c\vx\s\in
\I{C\vv}_{\eta,\s}= I_C(\vv\s,\vS)$ where $\vS=\I{\vv}_{\eta,\s}$. By
induction hypothesis, we have $\vv\s\in\SN$. So, let
$f:(\vz:\vV)(z:C\vz)(\vy:\vU)V$ be a recursor of $C$, $\vy\xi$ and
$\vy\t$ such that
$\xi_\vz^\vS,\t_\vz^{\vv\s}{}_z^{c\vx\s}\models\vy:\vU$. We must prove
that $v=f\vv\s^*(c\vx\s)\vy\t\in
S=\I{V}_{\xi_\vz^\vS,\t_\vz^{\vv\s}{}_z^{c\vx\s}}$. Since $v$ is
neutral, it suffices to prove that $\a\!\!(v)\sle S$. Since
$\vy\t\in\SN$, we can proceed by induction on $\vy\t$ with $\a$ as
well-founded ordering.

In the case of a reduction in $\vy\t$, we conclude by induction
hypothesis. In the case of a head-reduction, we conclude by
head-computability of $f$. And, in the case of a reduction in
$c\vx\s$, we conclude by the computability lemmas for function symbols
in \cite{blanqui01thesis}: if the strong normalization conditions are
satisfied and accessibility is correct w.r.t. computability, then
every reduct of $c\vx\s$ belongs to $\I{C\vv}_{\eta,\s}$. The fact
that accessibility is correct w.r.t. computability follows from the
completeness of the set of recursors w.r.t. accessibility.\qed
\end{proof}




\section{The Calculus of Inductive Constructions}

As an example, we prove the admissibility of a large class of weak
recursors for strictly positive types, from which Coq's recursors
\cite{coq02} can be easily derived. This can be extended to strong
recursors and to some non-strictly positive types (see Section
\ref{sec-pos}).

\begin{definition}
Let $C:(\vz:\vV)\st$ and $\vc$ be {\em strictly positive} constructors
of $C$, that is, if $c_i$ is of type $(\vx:\vT)C\vv$ then either no
type equivalent to $C$ occurs in $T_j$ or $T_j$ is of the form
$(\vec\alpha:\vW)C\vw$ with no type equivalent to $C$ occurring in
$\vW$. The {\em parameters} of $C$ is the biggest sequence $\vq$ such
that $C:(\vq:\vQ)(\vz:\vV)\st$ and each $c_i$ is of type
$(\vq:\vQ)(\vx:\vT)C\vq\vv$ with $T_j= (\vec\alpha:\vW)C\vq\vw$ if $C$
occurs in $T_j$.

The {\em canonical} weak recursor\footnote{Strong recursors cannot be
defined by taking $P:(\vz:\vV)C\vq\vz\A\B$ instead since
$(\vz:\vV)C\vq\vz\A\B$ is not typable in CC. They must be defined for
each $P$.} of $C$ w.r.t $\vc$ is $rec^*_{\vc}:
(\vq:\vQ)(\vz:\vV)(z:C\vq\vz)(P:(\vz:\vV)C\vq\vz\A\st)(\vy:\vU)P\vz z$
with $U_i= (\vx:\vT)(\vx':\vT')P\vv(c_i\vq\vx)$, $T_j'=
(\vec\alpha:\vW)P\vw(x_j\vec\alpha)$ if $T_j=
(\vec\alpha:\vW)C\vq\vw$, and $T_j'=T_j$ otherwise, defined by the
rules $rec^*_{\vc}\vq\vz(c_i\vq'\vx)P\vy\a y_i\vx\vt'$ where $t_j'=
[\vec\alpha:\vW](rec^*_{\vc}\vq\vw(x_j\vec\alpha)P\vy)$ if $T_j=
(\vec\alpha:\vW)C\vq\vw$, and $t_j'=x_j$ otherwise.\footnote{We could
erase the useless arguments $t_j'=x_j$ when $T_j'=T_j$.}
\end{definition}


\begin{lemma}
The set of canonical recursors is complete
w.r.t. accessibility.\footnote{In \cite{werner94thesis} (Lemma 4.35),
Werner proves a similar result.}
\end{lemma}

\begin{proof}
Let $c=c_i:(\vq:\vQ)(\vx:\vT)C\vq\vv$ be a constructor of
$C:(\vq:\vQ)(\vz:\vV)\st$, $\vq\eta$, $\vx\eta$, $\vq\s$ and $\vx\s$
such that $\vq\s\vv\s\in\SN$ and $c\vq\s\vx\s\in
\I{C\vq\vv}_{\eta,\s}= I_C(\vq\s\vv\s,\vq\xi\I{\vv}_{\eta,\s})$. Let
$\va=\vq\vx$ and $\vA=\vQ\vT$. We must prove that, for all $j$,
$a_j\s\in \I{A_j}_{\eta,\s}$. For the sake of simplicity, we assume
that weak and strong recursors have the same syntax. Since
$\vq\s\vv\s$ have normal forms, it suffices to find $u_j$ such that
$rec_c\vq\vv(c\vq\vx)P_j u_j\a u_j\vx\vt'\ab^* a_j$. Take $u_j=
[\vx:\vT][\vx':\vT']a_j$.\qed
\end{proof}


\begin{lemma}
Canonical recursors are head-computable.
\end{lemma}

\begin{proof}
Let
$f:(\vq:\vQ)(\vz:\vV)(z:C\vq\vz)(P:(\vz:\vV)C\vq\vz\A\st)(\vy:\vU)P\vz
z$ be the canonical weak recursor w.r.t. $\vc$,
$T=(\vz:\vV)C\vq\vz\A\st$, $c=c_i:(\vq:\vQ)(\vx:\vT)C\vq\vv$,
$\vq\eta$, $\vq\s$, $\vx\eta$, $\vx\s$, $P\xi$, $P\t$, $\vy\xi$,
$\vy\t$, $\vR=\I{\vv}_{\eta,\s}$, $\xi'=\xi_\vz^\vR$ and
$\t'=\t_\vz^{\vv\s}{}_z^{c\vx\s}$, and assume that
$\eta,\s\models\G_c$ and $\eta\xi',\s\t'\models P:T,\vy:\vU$. We must
prove that $y_i\t\vx\s\vt'\s\t\in \I{P\vz z}_{\xi',\t'}$.

We have $y_i\t\in \I{U_i}_{\xi',\t'}$, $U_i=
(\vx:\vT)(\vx':\vT')P\vv(c\vq\vx)$ and $x_j\s\in \I{T_j}_{\eta,\s}=
\I{T_j}_{\eta\xi',\s\t'}$. We prove that $t'_j\s\t\in
\I{T_j'}_{\eta\xi',\s\t'}$. If $T_j'=T_j$ then $t_j'\s\t=x_j\s$ and we
are done. Otherwise, $T_j= (\vec\alpha:\vW)C\vq\vw$, $T_j'=
(\vec\alpha:\vW)P\vw(x_j\vec\alpha)$ and $t_j'=
[\vec\alpha:\vW]f\vq\vw(x_j\vec\alpha)P\vy$. Let $\vec\alpha\zeta$ and
$\vec\alpha\g$ such that $\eta\xi'\zeta,\s\t'\g\models
\vec\alpha:\vW$. Let $t=x_j\s\vec\alpha\g$. We must prove that
$v=f\vq\s\vw\s\g tP\t\vy\t\in
S=\I{P\vw(x_j\vec\alpha)}_{\eta\xi'\zeta,\s\t'\g}$. Since $v$ is
neutral, it suffices to prove that $\a\!\!(v)\sle S$.

We proceed by induction on $\vq\s\vw\s\g tP\t\vy\t\in\SN$ with $\a$ as
well-founded ordering (we can assume that $\vw\s\g\in\SN$ since
$\th_f^<\tf:s_f$). In the case of a reduction in $\vq\s\vw\s\g
tP\t\vy\t$, we conclude by induction hypothesis. Assume now that we
have a head-reduct $v'$. By definition of recursors, $v'$ is also a
head-reduct of $v_0=f\vq\s^*\vw\s\g^*t P\t\vy\t$ where
$\vq\s^*\vw\s\g^*$ are the normal forms of $\vq\s\vw\s\g$. If $v_0\in
S$ then, by (R2), $v'\in S$. So, let us prove that $v_0\in S$.

By candidate substitution, $S=\I{P\vz
z}_{\xi_\vz^\vS,\t_\vz^{\vw\s\g}{}_z^t}$ with
$\vS=\I{\vw}_{\eta\xi'\zeta,\s\t'\g}= \I{\vw}_{\eta\xi\zeta,\s\t\g}$
for $\FV(\vw)\sle\{\vq,P,\vx,\vec\alpha\}$. Since $x_j\s\in
\I{T_j}_{\eta\xi',\s\t'}$ and $\eta\xi'\zeta,\s\t'\g\models
\vec\alpha:\vW$, $t\in \I{C\vq\vw}_{\eta\xi'\zeta,\s\t'\g}=
I_C(\vq\s\vw\s\g,\vq\xi\vS)$. Since $\eta\xi',\s\t'\models
P:T,\vy:\vU$ and $\FV(T\vU)\sle\{\vq,P\}$, we have
$\eta\xi,\s\t\models P:T,\vy:\vU$ and
$\eta\xi_\vz^\vS,\s\t_\vz^{\vw\s\g}{}_z^t\models
P:T,\vy:\vU$. Therefore, $v_0\in S$.\qed
\end{proof}


It follows that CAC essentially subsumes CIC as defined in
\cite{werner94thesis}. Theorem \ref{thm-admis} cannot be applied to
CIC directly since CIC and CAC do not have the same syntax and the
same typing rules. So, in \cite{blanqui01thesis}, we defined a
sub-system of CIC, called CIC$^-$, whose terms can be translated into
a CAC. Without requiring inductive types to be {\em safe} and to
satisfy (I6), we think that CIC$^-$ is essentially as powerful as CIC.

\begin{theorem}
The system CIC$^-$ defined in \cite{blanqui01thesis} (Chapter 7) is
strongly normalizing even though inductive types are unsafe and do not
satisfy (I6).
\end{theorem}




\section{Non-strictly positive types}
\label{sec-pos}

We are going to see that the use of elimination-based interpretations
allows us to have functions defined by recursion on non-strictly
positive types too, while CIC has always been restricted to strictly
positive types. An interesting example is given by Abel's
formalization of first-order terms with continuations as an inductive
type $trm:\st$ with the constructors \cite{abel02tr}:

\begin{typc}
var & nat\A trm\\
fun & nat\A (list~trm)\A trm\\
mu & \non\non trm\A trm\\
\end{typc}

\noindent
where $list:\st\A\st$ is the type of polymorphic lists, $\neg X$ is an
abbreviation for $X\A\bot$ (in the next section, we prove that $\neg$
can be defined as a function), and $\bot:\st$ is the empty type. Its
recursor $rec:(A:\st)(y_1:nat\A A)$ $(y_2:nat\A list~trm\A listA\A
A)(y_3:\neg\neg trm\A \neg\neg A\A A)(z:trm)A$ can be defined by:

\begin{rewc}
rec~A~y_1~y_2~y_3~(var~n) & y_1~n\\
rec~A~y_1~y_2~y_3~(fun~n~l) & y_2~n~l~(map~trm~A~(rec~A~y_1~y_2~y_3)~l)\\
rec~A~y_1~y_2~y_3~(mu~f) &
y_3~f~[x:\neg A](f~[y:trm](x~(rec~A~y_1~y_2~y_3~y)))\\
\end{rewc}

\noindent
where $map: (A:\st)(B:\st)(A\A B)\A list~A\A list~B$ is defined by:

\begin{rewc}
map~A~B~f~(nil~A') & (nil~B)\\
map~A~B~f~(cons~A'~x~l) & cons~B~(f~x)~(map~A~B~f~l)\\
map~A~B~f~(app~A'~l~l') & app~B~(map~A~B~f~l)~(map~A~B~f~l')\\
\end{rewc}


We now check that $rec$ is an admissible recursor. Completeness
w.r.t. accessibility is easy. For the head-computability, we only
detail the case of $mu$. Let $f\s$, $t=mu~f\s$, $A\xi$, $A\t$ and
$\vy\t$ such that $\vide,\s\models\G_{mu}$ and $\xi,\s\t_z^t\models
\G=A:\st$, $\vy:\vU$ where $U_i$ is the type of $y_i$. Let
$b=recA\t\vy\t$, $c=[y:trm](x(by))$ and $a=[x:\neg A\t](f\s c)$. We
must prove that $y_3\t f\s a\in \I{A}_{\xi,\s\t_z^t}=A\xi$.

Since $\xi,\s\t_z^t\models \G$, $y_3\t\in \I{\neg\neg trm\A \neg\neg
A\A A}_{\xi,\t}$. Since $\vide,\s\models\G_{mu}$, $f\s\in \I{\neg\neg
trm}$. Thus, we are left to prove that $a\in \I{\neg\neg A}_{\xi,\t}$,
that is, $f\s c\g\in I_\bot$ for all $x\g\in \I{\neg
A}_{\xi,\t}$. Since $f\s\in \I{\neg\neg trm}$, it suffices to prove
that $c\g\in \I{\neg trm}$, that is, $x\g(by\g)\in I_\bot$ for all
$y\g\in I_{trm}$. This follows from the facts that $x\g\in \I{\neg
A}_{\xi,\t}$ and $by\g\in A\xi$ since $y\g\in I_{trm}$.




\section{Inductive-recursive types}
\label{sec-indrec}

In this section, we define new positivity conditions for dealing with
{\em inductive-recursive type definitions} \cite{dybjer00jsl}. An
inductive-recursive type $C$ has constructors whose arguments have a
type $Ft$ with $F$ defined by recursion on $t:C$, that is, a predicate
$F$ and its domain $C$ are defined at the same time.

A simple example is the type $dlist:(A:\st)(\#:A\A A\A\st)\st$ of
lists made of distinct elements thanks to the predicate $fresh:
(A:\st)(\#:A\A A\A\st) A\A (dlist\,A\,\#)\A\st$ parametrized by a
function $\#$ to test whether two elements are distinct. The
constructors of $dlist$ are:

\begin{typc}
nil & (A:\st)(\#\!:\!A\!\A\! A\!\A\!\st)(dlist\,A\,\#)\\
cons & (A:\st)(\#\!:\!A\!\A\! A\!\A\!\st)
(x:A)(l:dlist\,A\,\#)(fresh~A~\#~x~l)\A (dlist\,A\,\#)\\
\end{typc}

\noindent
and the rules defining $fresh$ are:

\begin{rewc}
fresh~A~\#~x~(nil~A') & \top\\
fresh~A~\#~x~(cons~A'~y~l~h) & x\#y \et fresh~A~\#~x~l\\
\end{rewc}

\noindent
where $\top$ is the proposition always true and $\et$ the connector
``and''. Other examples are given by Martin-L\"of's definition of the
first universe {\em \`a la} Tarski \cite{dybjer00jsl} or by Pollack's
formalization of record types with manifest fields \cite{pollack02fac}.


\newcommand{\mon}{\mr{Mon}}

\begin{definition}[Positive/negative positions]
Assume that every predicate symbol $f:(\vx:\vt)U$ is equipped with a
set $\mon^+(f)\sle \{i\le \at_f~|~ x_i\in\XB\}$ of {\em monotone
arguments} and a set $\mon^-(f)\sle \{i\le \at_f~|~ x_i\in\XB\}$ of
{\em anti-monotone arguments}. The sets of {\em positive positions}
$\pos^+(t)$ and {\em negative positions} $\pos^-(t)$ in a term $t$ are
inductively defined as follows:

\begin{lst}{--}
\item $\pos^\d(s)= \pos^\d(x)= \{\vep~|~\d=+\}$,
\item $\pos^\d((x:U)V)= 1.\pos^{-\d}(U)\cup 2.\pos^\d(V)$,
\item $\pos^\d([x:U]v)= 2.\pos^\d(v)$,
\item $\pos^\d(tu)= 1.\pos^\d(t)$ if $t\neq f\vt$,
\item $\pos^\d(f\vt)= \{1^{|\vt|}~|~\d=+\}\cup
  \,\bigcup\{1^{|\vt|-i}2.\pos^{\ep\d}(t_i)~|~ \ep\in\{-,+\},\,
  i\in\mon^\ep(f)\}$,
\end{lst}

\noindent
where $\d\in\{-,+\}$, $-+=-$ and $--=+$ (usual rule of signs).
\end{definition}


\begin{theorem}[Strong normalization]
Definition \ref{def-rec} is modified as follows. A pre-recursor
$f:(\vz:\vV)(z:C\vz)W$ is a {\em recursor} if:

\begin{lst}{\bu}
\item no $F>C$ occurs in $W$,
\item every $F\simeq C$ occurs only positively in $W$,
\item if $i\in\mon^\d(C)$ then $\pos(z_i,W)\sle \pos^\d(W)$.
\end{lst}

\noindent
Assume furthermore that, for every rule $F\vl\a r$:

\begin{lst}{\bu}
\item no $G>F$ occurs in $r$,
\item for all $i\in\mon^\ep(F)$, $l_i\in\XB$ and $\pos(l_i,r)\sle
  \pos^\ep(r)$.
\end{lst}

\noindent
Then, Theorem \ref{thm-admis} is still valid.
\end{theorem}

\begin{proof}
For Theorem \ref{thm-admis} to be still valid, we must make sure that
$\vphi$ (see Definition \ref{def-int}) is still monotone, hence has a
least fixpoint. To this end, we need to prove that $\I{t}_{\xi,\t}^I$
is monotone (resp. anti-monotone) w.r.t. $x\xi$ if $x$ occurs only
positively (resp. negatively) in $t$, and that $\I{t}_{\xi,\t}^I$ is
monotone (resp. anti-monotone) w.r.t. $I_C$ if $C$ occurs only
positively (resp. negatively) in $t$. These results are easily
extended to the new positivity conditions by reasoning by induction on
the well-founded ordering used for defining the defined predicate
symbols.

Let us see what happens in the case where $t=F\vt$ with $F$ a defined
predicate symbol. Let $\le^+=\le$ and $\le^-=\ge$. We want to prove
that, if $\xi_1\le_x \xi_2$ ({\em i.e.} $x\xi_1\le x\xi_2$ and, for
all $y\neq x$, $y\xi_1=y\xi_2$) and $\pos(x,t)\sle \pos^\d(t)$, then
$\I{t}^I_{\xi_1,\t}\le^\d \I{t}^I_{\xi_2,\t}$. By definition of $I_F$,
if the normal forms of $\vt\t$ matches the left hand-side of $F\vl\a
r$, then $\I{F\vt}^I_{\xi_i,\t}= \I{r}^I_{\xi_i',\s}$ where $\s$ is
the matching substitution and, for all $y\in\FV(r)$, $y\xi_i'=
\I{t_{\ka_y}}^I_{\xi_i,\t}$ where $\ka_y$ is such that $l_{\ka_y}=y$
(see \cite{blanqui01thesis} for details). Now, since $\pos(x,F\vt)\sle
\pos^\d(F\vt)$, $\pos(x,t_{\ka_y})\sle \pos^{\ep\d}(t_{\ka_y})$ for
some $\ep$. Hence, by induction hypothesis, $\xi_1'\le_y^{\ep\d}
\xi_2'$. Now, since $\pos(y,r)\sle \pos^\ep(r)$, by induction
hypothesis again, $\I{r}_{\xi_1',\s}\le^{\ep^2\d}=\le^\d
\I{r}_{\xi_2',\s}$.\qed
\end{proof}

For instance, in the positive type $trm$ of Section \ref{sec-pos},
instead of considering $\neg\neg A$ as an abbreviation, one can
consider $\neg$ as a predicate symbol defined by the rule $\neg A\a
A\A\bot$ with $\mon^-(\neg)=\{1\}$. Then, one easily checks that $A$
occurs negatively in $A\A\bot$, and hence that $trm$ occurs positively
in $\neg\neg trm$ since $\pos^+(\neg\neg trm)= \{1\}\cup 2.\pos^-(\neg
trm)= \{1\}\cup 2.2.\pos^+(trm)= \{1,2.2\}$.


\section{Conclusion}

By using an elimination-based interpretation for inductive types, we
proved that the Calculus of Algebraic Constructions completely
subsumes the Calculus of Inductive Constructions. We define general
conditions on extended recursors for preserving strong normalization
and show that these conditions are satisfied by a large class of
recursors for strictly positive types and by non-strictly positive
types too. Finally, we give general positivity conditions for dealing
with inductive-recursive types.\\

\small\noindent{\bf Acknowledgments.} I would like to thank C. Paulin,
R. Matthes, J.-P. Jouannaud, D. Walukiewicz, G. Dowek and the
anonymous referees for their useful comments on previous versions of
this paper. Part of this work was performed during my stay at
Cambridge (UK) thanks to a grant from the INRIA.


\begin{thebibliography}{10}

\bibitem{abel02tr}
A.~Abel.
\newblock Termination checking with types.
\newblock Technical Report 0201, Ludwig Maximilians Universit\"at, M\"unchen,
  Germany, 2002.

\bibitem{barbanera94lics}
F.~Barbanera, M.~Fern\'andez, and H.~Geuvers.
\newblock Modularity of strong normalization and confluence in the
  algebraic-$\lambda$-cube.
\newblock In {\em Proceedings of the 9th IEEE Symposium on Logic in Computer
  Science\em, 1994}.

\bibitem{barendregt92book}
H.~Barendregt.
\newblock Lambda calculi with types.
\newblock In S.~Abramski, D.~Gabbay, and T.~Maibaum, editors, {\em Handbook of
  logic in computer science}, volume~2. Oxford University Press, 1992.

\bibitem{blanqui01lics}
F.~Blanqui.
\newblock Definitions by rewriting in the {C}alculus of {C}onstructions
  (extended abstract).
\newblock In {\em Proceedings of the 16th IEEE Symposium on Logic in Computer
  Science\em, 2001}.

\bibitem{blanqui01thesis}
F.~Blanqui.
\newblock {\em Th\'eorie des {T}ypes et {R}\'ecriture}.
\newblock PhD thesis, Universit\'e Paris XI, Orsay, France, 2001.
\newblock Available in english as "Type Theory and Rewriting".

\bibitem{blanqui03cac}
F.~Blanqui.
\newblock Definitions by rewriting in the {C}alculus of {C}onstructions, 2003.
\newblock Journal submission, 68 pages.

\bibitem{blanqui03short}
F.~Blanqui.
\newblock A short and flexible strong normalization proof for the {C}alculus of
  {A}lgebraic {C}onstructions with curried rewriting, 2003.
\newblock Draft.

\bibitem{coquand92types}
T.~Coquand.
\newblock Pattern matching with dependent types.
\newblock In {\em Proceedings of the International Workshop on Types for Proofs
  and Programs, 1992}.
\newblock \url{http://www.lfcs.informatics.ed.ac.uk/research/types-bra/proc/}.

\bibitem{coquand88ic}
T.~Coquand and G.~Huet.
\newblock The {C}alculus of {C}onstructions.
\newblock {\em Information and Computation}, 76(2--3):95--120, 1988.

\bibitem{coquand88colog}
T.~Coquand and C.~Paulin-Mohring.
\newblock Inductively defined types.
\newblock In {\em Proceedings of the International Conference on Computer
  Logic\em, Lecture Notes in Computer Science 417, 1988}.

\bibitem{dershowitz90book}
N.~Dershowitz and J.-P. Jouannaud.
\newblock Rewrite systems.
\newblock In J.~van Leeuwen, editor, {\em Handbook of Theoretical Computer
  Science}, volume~B, chapter~6. North-Holland, 1990.

\bibitem{dybjer00jsl}
P.~Dybjer.
\newblock A general formulation of simultaneous inductive-recursive definitions
  in type theory.
\newblock {\em Journal of Symbolic Logic}, 65(2):525--549, 2000.

\bibitem{girard88book}
J.-Y. Girard, Y.~Lafont, and P.~Taylor.
\newblock {\em Proofs and Types}.
\newblock Cambridge University Press, 1988.

\bibitem{harper99ipl}
R.~Harper and J.~Mitchell.
\newblock Parametricity and variants of {G}irard's {J} operator.
\newblock {\em Information Processing Letters}, 70:1--5, 1999.

\bibitem{jouannaud91lics}
J.-P. Jouannaud and M.~Okada.
\newblock Executable higher-order algebraic specification languages.
\newblock In {\em Proceedings of the 6th IEEE Symposium on Logic in Computer
  Science\em, 1991}.

\bibitem{klop93tcs}
J.~W. Klop, V.~van Oostrom, and F.~van Raamsdonk.
\newblock Combinatory reduction systems: introduction and survey.
\newblock {\em Theoretical Computer Science}, 121:279--308, 1993.

\bibitem{matthes98thesis}
R.~Matthes.
\newblock {\em Extensions of System F by Iteration and Primitive Recursion on
  Monotone Inductive Types}.
\newblock PhD thesis, Ludwig Maximilians Universit\"at, M\"unchen, Germany,
  1998.

\bibitem{mcbride99thesis}
C.~McBride.
\newblock {\em Dependently typed functional programs and their proofs}.
\newblock PhD thesis, University of Edinburgh, United Kingdom, 1999.

\bibitem{mendler87thesis}
N.~P. Mendler.
\newblock {\em Inductive Definition in Type Theory}.
\newblock PhD thesis, Cornell University, United States, 1987.

\bibitem{paulin01pc}
C.~Paulin-Mohring.
\newblock Personal communication, 2001.

\bibitem{pollack02fac}
R.~Pollack.
\newblock Dependently typed records in type theory.
\newblock {\em Formal Aspects of Computing}, 13(3--5):341--363, 2002.

\bibitem{coq02}
Coq~Development Team.
\newblock {\em The {C}oq Proof Assistant Reference Manual -- Version 7.3}.
\newblock INRIA Rocquencourt, France, 2002.
\newblock \url{http://coq.inria.fr/}.

\bibitem{werner94thesis}
B.~Werner.
\newblock {\em Une Th\'eorie des {C}onstructions {I}nductives}.
\newblock PhD thesis, Universit\'e Paris VII, France, 1994.

\end{thebibliography}

\end{document}